\documentclass[12pt]{article}
\usepackage{english}
\begin{document}                                                                
\thispagestyle{empty}                                                           
\bigskip\bigskip                                                                
\begin{center}                                                                  
{\bf \Large{Comment on ``Percolation Properties of the 2D Heisenberg 
Model''}}
\end{center}                                                                    
\vskip 1.0truecm                                                                
\centerline{\bf                                                                 
Adrian Patrascioiu}                                                             
\vskip5mm                                                                       
\centerline{Physics Department,                                                 
University of Arizona, Tucson, AZ 85721, U.S.A.}                                
\vskip5mm                                                                       
\centerline{and}                                                                
\vskip5mm                                                                       
\centerline{\bf Erhard Seiler}                                                  
\vskip5mm                                                                       
\centerline{Max-Planck-Institut f\"ur                                           
 Physik}                                                                        
\centerline{ -- Werner-Heisenberg-Institut -- }                                 
\centerline{F\"ohringer Ring 6, 80805 Munich, Germany}                          
\vskip 2cm                                                                      
\begin{abstract}
We comment on a recent paper by All\`es et al \cite{aacp}.
\end{abstract}
                                                                                
\newcommand{\bee}{\begin{equation}}                                             
\newcommand{\ee}{\end{equation}}                                                
\newcommand{\ba}{\begin{array}}                                                 
\newcommand{\ea}{\end{array}}                                                   
\newcommand{\bea}{\begin{eqnarray}}                                             
\newcommand{\eea}{\end{eqnarray}}                                               
\newcommand{\rmD}{\mbox{\scriptsize D}}                                         
\newcommand{\rmSI}{\mbox{\scriptsize SI}}                                       
\newcommand{\RR}{{\Bbb R}}                                                      
In a recent letter All\`es at al \cite{aacp} claim to show that the two
dimensional classical Heisenberg model does not have a massless phase. The
paper is an attempt to refute certain arguments advanced by us in 1991
\cite{ap}
\cite{pat,ps}, which remain posted among the {\it Open Problems in
Mathematical Physics}  on the web site of the International Association
of Mathematical Physics \cite{IAMP}; thus the claim of Alles et al,
if correct, would be very important. We appreciate that the authors
made an effort to falsify our arguments, but in fact the paper is far
from establishing its claim. It shows some fundamental misunderstanding
of our arguments and contains several incorrect statements.

1. To simplify the discussion, following All\`es et al, we will ignore 
the distinction between $\ast$-percolation and percolation; as
suggested by their choice of cluster, we will consider the model with
standard nearest neighbor action at inverse temperature $\beta$ as a
model with a Lipschitz constraint $|s-s'|<\epsilon$ for neighboring
spins $s$,$s'$ and suitable $\epsilon$. Then from our 1991
arguments it follows rigorously that as long as the FK-clusters have 
finite mean size, i.e. in the massive phase, the equatorial cluster  
of width $\epsilon$ must percolate. Everybody agrees that at
$\beta=2.0$ the standard action model has a finite correlation
length $\xi$ and in fact All\`es et al even state the value of $L/\xi$. 
Therefore their finding that the equatorial cluster they considered 
percolates is nothing but another indication that at $\beta=2.0$ the 
model is still massive. To prove that our 1991 conjecture is false, 
they would have to show that for any arbitrarily small $\epsilon$ 
there exists a finite $\beta_p(\epsilon)$ such that for any 
$\beta>\beta_p(\epsilon)$ the equatorial cluster $S_\epsilon$
percolates.

2. In 1991 we also gave an auxiliary argument: if, contrary to our
expectation, an arbitrarily thin equatorial cluster percolated at 
sufficiently
large $\beta$, then the $O(3)$ symmetry would be broken, since then
there would be a much larger equatorial cluster on which the induced 
$O(2)$ model would be in its massless (KT) phase. All\`es et all suggest
that this argument fails because the percolating cluster they found at 
$\beta=2.0$ is very flimsy, having a fractal dimension less than 2. 
To support their reasoning, they quote a result of Koma and Tasaki 
\cite{kt} that for $D<2$ the KT phase does not exist. The claim that 
this percolating cluster has a fractal dimension less than 2 is in 
conflict with their own numbers in Tab.1, showing that this cluster 
has a nonvanishing density. This is of course not surprising, since 
for a translation invariant percolation problem in $2D$ the (unique) 
percolating cluster under rather general conditions always has a 
finite density \cite{newman}. 

3. In support of the fractal picture, All\`es et al point out the
fact that the ratio of the perimeter over the area of the cluster does
not go to 0 as its size increases. Consider then a regular square lattice 
on which square holes of size $L\times L$ have been made, in such a way 
that a percolating subset remains. The ratio of its perimeter to its
area does not vanish and there should be no doubt that on such a lattice 
the $O(2)$ model has a KT phase for any finite $L$.

4. In view of this, we think that even on the `flimsy' percolating 
cluster found by All\`es et all, the $O(2)$ model does have a KT phase
at low enough temperature. It would be interesting to verify this,
even though our argument does not depend on the
existence of such a transition on that particular percolating 
cluster.
\medskip\noindent

\end{document}